\documentclass{aa} 
\usepackage[varg]{txfonts}
\usepackage{graphicx}

\begin{document}

\title{Efficient injection from large telescopes into single-mode fibres: Enabling the era of ultra-precision astronomy}

\author{N. Jovanovic\inst{1,2}\thanks{\email{jovanovic.nem@gmail.com}}, C. Schwab\inst{2,3}, O. Guyon\inst{1,4,5,6}, J. Lozi\inst{1}, N. Cvetojevic\inst{3,7,8}, F. Martinache\inst{9},\\ S. Leon-Saval\inst{7}, B. Norris\inst{7}, S. Gross\inst{2,8}, D. Doughty\inst{1}, T. Currie\inst{1} and N. Takato\inst{1}}

\institute{Subaru Telescope, National Astronomical Observatory of Japan, National Institutes of Natural Sciences (NINS), 650 North A'Ohoku Place, Hilo, HI, 96720, U.S.A.
\and MQ Photonics Research Centre, Department of Physics and Astronomy, Macquarie University, NSW 2109, Australia
\and Australian Astronomical Observatory, 105 Delhi Rd, North Ryde NSW 2113, Australia
\and Steward Observatory, University of Arizona, Tucson, AZ, 85721, U.S.A.
\and College of Optical Sciences, University of Arizona, Tucson, AZ 85721, U.S.A.
\and Astrobiology Center of NINS, 2-21-1, Osawa, Mitaka, Tokyo, 181-8588, Japan
\and Sydney Institute for Astronomy (SIfA), Institute for Photonics and Optical Science (IPOS), School of Physics, University of Sydney, NSW 2006, Australia
\and Centre for Ultrahigh-bandwidth Devices for Optical Systems (CUDOS)
\and Laboratoire Lagrange, Universit\'{e} C\^{o}te d’Azur , Observatoire de la C\^{o}te d’Azur, CNRS, Parc Valrose, Bât. H. FIZEAU, 06108 Nice, France}

\abstract{Photonic technologies offer numerous advantages for astronomical instruments such as spectrographs and interferometers owing to their small footprints and diverse range of functionalities. Operating at the diffraction-limit, it is notoriously difficult to efficiently couple such devices  directly with large telescopes. We demonstrate that with careful control of both the non-ideal pupil geometry of a telescope and residual wavefront errors, efficient coupling with single-mode devices can indeed be realised. A fibre injection was built within the Subaru Coronagraphic Extreme Adaptive Optics (SCExAO) instrument. Light was coupled into a single-mode fibre operating in the near-IR (J-H bands) which was downstream of the extreme adaptive optics system and the pupil apodising optics. A coupling efficiency of $86\%$ of the theoretical maximum limit was achieved at $1550$~nm for a diffraction-limited beam in the laboratory, and was linearly correlated with Strehl ratio. The coupling efficiency was constant to within $<30\%$ in the range $1250$--$1600$~nm. Preliminary on-sky data with a Strehl ratio of $60\%$ in the H-band produced a coupling efficiency into a single-mode fibre of $\sim50\%$, consistent with expectations. The coupling was $>40\%$ for $84\%$ of the time and $>50\%$ for $41\%$ of the time. The laboratory results allow us to forecast that extreme adaptive optics levels of correction (Strehl ratio $>90\%$ in H-band) would allow coupling of $>67\%$ (of the order of coupling to multimode fibres currently) while standard levels of wavefront correction (Strehl ratio $>20\%$ in H-band) would allow coupling of $>18\%$. For Strehl ratios $<20\%$, few-port photonic lanterns become a superior choice but the signal-to-noise, and pixel availability must be considered. These results illustrate a clear path to efficient on-sky coupling into a single-mode fibre which could be used to realise modal-noise-free radial velocity machines, very-long-baseline optical/near-IR interferometers and/or simply exploit photonic technologies in future instrument design.}


\titlerunning{Efficient single-mode fibre injection}
\authorrunning{N. Jovanovic et. al.}
\maketitle

\section{Introduction}
The key to precision astronomical measurements is the ability to carefully calibrate the data. The quality of calibration would be greatly improved if it were possible to operate at the diffraction-limit. In this case we mean that there is no wavefront error (i.e. a perfectly flat wavefront) and the point-spread function (PSF) is temporally invariant (stable with time). In the field of high-contrast imaging for example, this would enable a perfect PSF subtraction, revealing structures at unprecedented contrast around the host star, such as an exoplanet or disk. In the area of precision radial velocity, one of the key limitations is the stability of the slit (or fibre) illumination~\citep{cha2012}. A temporally invariant, diffraction-limited PSF would eliminate this issue and result in a consistently high level of performance by virtue of improved calibration for the data. Finally, interferometers would deliver stable fringes, enhancing the contrast with which faint structures around a host star could be studied. This would result in a more precise determination of the objects' properties~\citep{Coude1994}.

However, ground-based observations are hampered by atmospheric turbulence. One solution is to operate the instrument in space, free from the atmosphere. This comes at great cost and complexity. More commonly, adaptive optics (AO) systems are used upstream of the instrument. AO systems are based on measuring the incident wavefront and subsequently correcting for it with a deformable mirror capable of operating at high speed ($100$s of Hz). Such instruments are now ubiquitous at large ($>4$~m-class) observatories~\citep{Wiz2000,Minowa2010}. These systems offer Strehl ratios of $20-50\%$ in the H-band in median seeing. This translates to $200$-$300$~nm RMS residual wavefront error, which presents an improvement with which data can be calibrated, but is far from the ideal scenario. 

Recently, several large observatories have commissioned so-called "extreme" AO (ExAO) systems capable of delivering $90\%$ Strehl ratios in better than median seeing in the H-band~\citep{Dek2013,Mac2014,Vigan2016}. These systems reduce the residual wavefront error down to $\sim80$~nm RMS in H-band by controlling a larger region around the PSF~\citep{opp2003} in the focal plane and doing so at higher speed than conventional AO systems (refer to~\cite{jov2016b} for full details). Although these systems do not deliver the perfectly flat wavefront desired, they certainly approach this limit and improve the quality of calibrated astronomical data. 

To flatten the wavefront post-ExAO correction to the ideal case (i.e. $0$~nm RMS), some sort of spatial filtering is needed. This can be achieved in one of two ways: The light can be passed through a pinhole or a single-mode fibre (SMF)~\citep{Coude1994}. A SMF is named as such because it can only transport light through the fibre in one state, the fundamental mode. This mode has a near-Gaussian intensity profile and a flat phase front (to be precise, there are two identical modes, which have orthogonal polarisations). When light is passed through such a fibre, it exits with these properties imprinted on the beam. Most importantly however, the output intensity profile of a SMF is temporally invariant. This makes a SMF the ideal way to feed light to an astronomical instrument for precision measurements. Indeed, a discussion highlighting the impact of a SMF feed on stellar spectroscopy was presented by~\cite{crepp14} and more recently treated in detail by~\cite{jov2016b}.     

Once the light is in the SMF, a suite of photonic technologies that operate at the diffraction-limit become accessible for instrument design. Photonic components, either optical-fibre-based or on a wafer, offer a diverse range of functionality including the possibility for spectral filtering~\citep{Marien2012,Trinh2013}, dispersion~\citep{Cvetojevic2009,Cvetojevic2012}, reformatting~\citep{Iza2013,Harris2015} and calibration~\citep{Feger2014,Hal2014,Schwab2015} to name a few. In addition, photonic devices are compact and robust and can easily be stabilised to a high degree. This makes these devices highly desirable for implementation in astronomical instrument design.  

However,  efficiently coupling a SMF to a large telescope is inherently difficult because of the need to match the intensity distribution and phase front of the incident beam to that of the fundamental mode of the SMF. Early work was conducted by~\cite{Coude1992} investigated injecting seeing-limited light from two $0.8$~m telescopes into SMFs optimised for the K-band for the purposes of conducting interferometric measurements with the FLUOR beam combiner. This was followed by~\cite{Coude2000} who demonstrated a coupling efficiency of $20$-$25\%$ in K-band into a fluoride-based SMF on the $3.6$~m La Silla Telescope behind the ADONIS AO system. This pioneering work achieved over $50\%$ of the theoretical maximum coupling ($39\%$) for the experiential setup used. It was primarily limited by the relatively large central obstruction ($43.6\%$ of the primary), quasi-static low-order aberrations and a $25$~Hz telescope vibration. Since this time there have been several other demonstrations of SMF injection for both spectroscopy~\citep{ghas2012} and interferometry~\citep{perrin2006,mennesson2010}. More recently, light has been coupled with AO  assistance from one of the LBT $8$-m mirrors to a SMF in the y-band~\citep{bechter2016}. Preliminary results from that work yielded a coupling efficiency in the range of $20$--$25\%$, which is high considering the relatively short wavelength.  

In the era of ExAO equipped telescopes, achieving high coupling efficiencies becomes a distinct possibility and forms the basis for this study. In this body of work, we demonstrate efficient coupling into a SMF ($>70\%$) is possible behind an ExAO system with the use of pupil apodisation optics for the first time. This work builds upon the recent results presented in~\cite{jov2014,jov2016a}. Section~\ref{sec:how} gives an overview of how to couple into SMFs supported by simulations. Section~\ref{laboratory} outlines the experimental setup used while Section~\ref{labresults} summarises the results of the detailed coupling efficiency study conducted in the laboratory, which includes a comparison with a few-port photonic lantern. Section~\ref{sec:onskyresults} highlights the preliminary on-sky results while Section~\ref{sec:discussion} goes into a deeper discussion of their implications.  Section~\ref{sec:summary} rounds off the paper with a summary of the key results.

\section{How to efficiently inject light into single-mode fibres}\label{sec:how}
\subsection{Requirements for optimum coupling}
As outlined in the introduction, a SMF guides light in the fundamental mode with a Gaussian intensity profile and a flat phase front. In order to maximise coupling of light into this mode, the incident beam should match these specifications closely. The mode can be characterised in a number of ways. The mode field diameter (MFD) specifies the size of the mode and is measured to the point where the intensity profile drops to $1/e^{2}$ of the peak value. The fibre also has a numerical aperture (NA) that stipulates the angles of rays that can be accepted by the fibre and is measured to the $1\%$ intensity point of the far-field emanating from the fibre. Carefully matching the properties of a Gaussian beam to the MFD and NA is important for SMFs. If we take the case of a typical near-IR optimised SMF such as SMF-$28$ (from Corning), it offers a MFD~$=10.4~\mu$m at $1550$~nm and a NA$=0.14$. To match the spot size, a focal ratio $f/\#=5.27$ Gaussian beam (measured to the $1/e^2$ point) would need to be used. Any deviations in the $f/\#$ would lead to a reduction in the coupling efficiency. This corresponds to a NA for the beam of $0.095$, measured to the $1/e^2$ points, or a NA of $0.15$ measured to the $1\%$ points (more typical for fibres), which is little above the cutoff of the fibre but very close. What is important though is that the spot size of a Gaussian beam is matched to the MFD for optimum performance. In cases where the input beam is not a Gaussian, the optimum beam parameters to maximise coupling need to be calculated using the overlap integral method described in the following section. Simply matching the beam sizes for two dissimilar profile beams is typically a good starting point, but some optimisation is required.         

\subsection{Considerations for optimising the PSF for efficient coupling} 
\begin{figure}
\centering 
\includegraphics[width=0.99\linewidth]{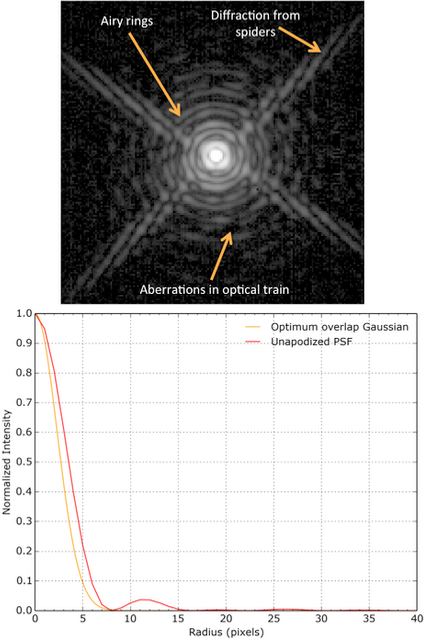}
\caption{\footnotesize (Top) Laboratory image of the PSF of the Subaru Telescope in the H-band. Various features which are caused by diffraction or aberrations are highlighted. (Bottom) Cross-sectional line profile of an Airy pattern and the optimum fitting Gaussian.}
\label{fig:aberrations}
\end{figure}

In order to determine how to optimise the coupling of light into a SMF, it is instrumental to first examine the PSF of a telescope. A telescope which is uniformly illuminated by starlight will form an Airy pattern in the focal plane. The PSF observed at Subaru Telescope is shown in the top panel of Fig.~\ref{fig:aberrations} as an example with a cross-sectional line profile offered in the bottom panel (red curve).
The diffraction rings around the PSF are the result of the fact that the pupil has a finite extent. The phase across the central core of the PSF is constant. Between each concentric ring is a dark region where the phase flips by $\pi$. The phase within each ring is constant and continues to flip for each successive ring. It is this constantly flipping phase that induces destructive interference when attempting to couple light from an Airy pattern into a Gaussian mode, which sets the upper limit. The coupling efficiency can be calculated by using the overlap integral of the two fields given by
\begin{equation}\label{eq:overlapint}
\eta= \frac{|\int_{}^{}E_{1}^{*}E_{2} dA|^{2}}  {\int_{}^{}|E_{1}|^{2}dA\int_{}^{}|E_{2}|^{2}dA},
\end{equation}
where $E_{1}$ and $E_{2}$ are the two complex fields (including amplitude and phase) that are to be overlapped~\citep{wag1982}. The maximum theoretical coupling efficiency of an unobstructed circular aperture into a SMF is $\sim80\%$~\citep{Shaklan1988}. An explanation of where this limit comes from is provided in section~\ref{couplimit}. The cross-sectional line profile of the optimum fit Gaussian to the Airy pattern presented in the bottom panel of Fig.~\ref{fig:aberrations}, is displayed as the yellow curve and was determined by evaluating Eq.~\ref{eq:overlapint}. 

The top panel in Fig.~\ref{fig:aberrations} also clearly shows strong diffraction from the telescope spiders. These push light away from the central peak of the Gaussian mode reducing coupling as well. Finally, quasi-static aberrations in the optical train, which manifest as asymmetries in the PSF core and Airy rings, indicate that the wavefront of the system is not perfectly flat, further reducing the coupling efficiency. 

The following subsections examine the effect of each of these in turn and outline ways to remedy them. 

\subsubsection{The central obstruction}
\begin{figure}[b!]
\centering 
\includegraphics[width=0.99\linewidth]{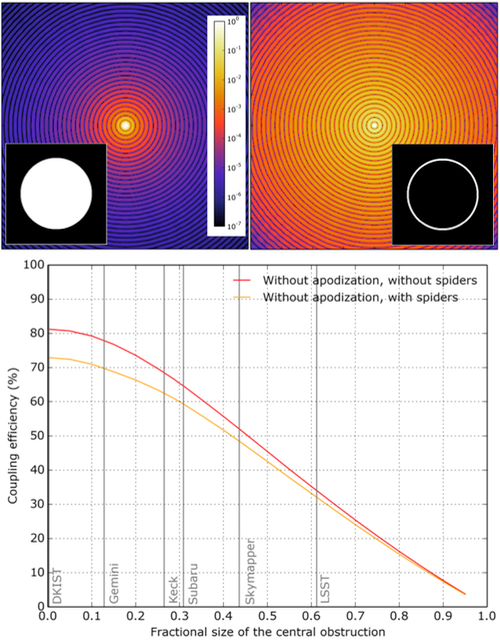}
\caption{\footnotesize Logarithmic stretch of a simulated PSF with (Top left) no central obstruction and (Top right) a relatively large central obstruction. Pupil geometries are shown in the insets. (Bottom) Coupling efficiency to a SMF from the PSF of a telescope as a function of the relative size of the central obstruction. The simulations were made with and without spiders. Grey vertical lines indicate the relative size of the secondary for several major observatories. We note that DKIST has a line at $0$.}
\label{fig:secondary}
\end{figure}

There are very few unobstructed telescope pupils at most large observatories. The presence of a central obstruction (obstructed circular aperture) moves power out of the central core of the PSF and into the rings (top panels of Fig.~\ref{fig:secondary}). The left image in the top panel shows a simulated PSF for an unobstructed pupil as compared to the right image, which is for an almost completely obstructed pupil. Both images are logarithmically stretched and it can be seen that successive diffraction rings have more similar peak surface brightness across the image in the case of the highly obstructed pupil. 

The bottom panel of Fig.~\ref{fig:secondary} shows the coupling efficiency into a SMF as a function of the size of the secondary obstruction (red curve).
Overlaid in the figure are lines that correspond to the size of the secondary obstruction for various well-known observatories. It can be seen that as the relative size of the secondary increases, the coupling efficiency reduces in a monotonic way which becomes quasi-linear beyond a $20\%$ relative size secondary. Since the optimum coupling efficiency is achieved by only including light from the PSF core, as explained above and in section~\ref{couplimit}, then by moving more light into the Airy rings, due to a larger secondary obstruction, the coupling efficiency is expected to reduce as seen in the simulation and was first described by~\cite{Coude1994}. 

DKIST~\citep{mat2016}, a telescope optimised for high-contrast imaging observations of the sun and the only off-axis telescope with no obstruction of those chosen, offers the greatest coupling ($\sim81\%$). The Gemini Observatory offers the next best coupling efficiency given their relatively small secondary obstruction ($\sim80\%$). Subaru Telescope has a maximum coupling efficiency of $65\%$, which is similar to Keck, while the lowest coupling efficiency is offered by the telescopes optimised for wide-field imaging, Skymapper and LSST. This chart demonstrates that it is important to consider the suitability of the observatory to SMF injection, as the final performance will depend on the geometry of the telescope used.

\subsubsection{Telescope spiders}
The yellow curve in Fig.~\ref{fig:secondary} (bottom) shows the same coupling efficiency as descrtibed above but this time taking into account the spiders at Subaru Telescope, where the experiments in this study were carried out. The coupling efficiency has been reduced by $\sim10\%$ across the board. This is a result of the light being diffracted into high-spatial-frequency features which form a cross-like pattern about the PSF and have a poor overlap with the mode of a fibre (top panel in Fig.~\ref{fig:aberrations}). The $10\%$ loss is higher than for other observatories as Subaru has a heavy primary focus unit that requires thicker supports and the ExAO system (SCExAO) utilises a fixed pupil mask inside the instrument, which has slightly oversized spiders for calibration purposes (spider thickness inside the ExAO system is $4.8\%$ of the size of the outer pupil diameter). Again, by carefully selecting a telescope with thinner secondary supports, it is possible to minimise this coupling loss. It should be made clear at this point that the $10\%$ shown in the figure is only due to the drop in coupling efficiency, and does not take into account the attenuation of the light due to the fact the spiders obscure a portion of the pupil. A method for removing the effect of the spiders was proposed and is outlined in section~\ref{spiders} for completeness.

\subsubsection{Quasi-static and turbulent aberrations}
Besides the diffractive effects described above, maintaining a flat wavefront is critical to maximising the coupling efficiency, especially when operating at a ground-based observatory. Aberrations that distort the wavefront can be divided into two main categories: Quasi-static and turbulent aberrations. Here we offer some qualitative insight into how each can be addressed to optimise the coupling into the fibre. 

Quasi-static aberrations are a result of defects in the optical train. These could be low-order aberrations, well represented by Zernike polynomials, which come from optical misalignment or low-quality optics. They can also be high-order aberrations induced by the fabrication process of the optics. They are termed quasi-static because they evolve slowly with time as various optical elements move to track the star (i.e. the telescope pointing, the image rotator, and so on). On the other hand, turbulent aberrations are the result of light propagating through the turbulent atmosphere. These are much faster and vary with the coherence time of the atmosphere ($\sim5$~ms for Maunakea). No matter which type of aberration is present, they both distort the flat wavefront from the star present at the top of the atmosphere, reducing the coupling into a SMF. Indeed, the coupling is correlated with the Strehl ratio and so if the cumulative wavefront error is large (and the Strehl ratio is low), then the coupling will also be low. We come back to this point in the experimental section.     

To compensate for these aberrations and restore the flat wavefront, wavefront control is required. An AO or ExAO system is critical to address the turbulent aberrations. For this reason we employ the pyramid wavefront sensor in the SCExAO instrument in this body of work~\citep{jov2015,jov2016b}. However, a pyramid wavefront sensor has no knowledge of what a flat wavefront looks like as it is a relative wavefront sensor and can only maintain the wavefront with respect to some reference. Therefore, in order to set the reference point of the wavefront sensor to a flat wavefront, and remove all quasi-static aberrations in the optical train, an absolute wavefront sensor must be used in addition to the pyramid wavefront sensor. There are numerous solutions to this but one wavefront sensor recently demonstrated on-sky that could be used for this is the asymmetric pupil Fourier wavefront sensor~\citep{Mar2016}. The sensor relies on the Fourier analysis of focal plane images with an asymmetric mask introduced into the pupil of the instrument. It allows for the aberrations in the pupil to be determined by examining the image in the focal plane, that is, where the SMF is located. It was tested both on and off sky on the SCExAO instrument operating in the H-band on the lowest $10$ or so Zernike modes and has demonstrated that it can indeed drive the wavefront to a flat solution, and so was used throughout this work. 

Regardless of which sensor is to be used, it is clear that a combination of wavefront sensors will be required to deal with both fast and slowly varying aberrations that have a range of amplitudes in order to provide a flat wavefront for injection into a fibre.

\subsection{Pupil apodisation}\label{sec:pupapo}
\begin{figure}[b!]
\centering 
\includegraphics[width=0.99\linewidth]{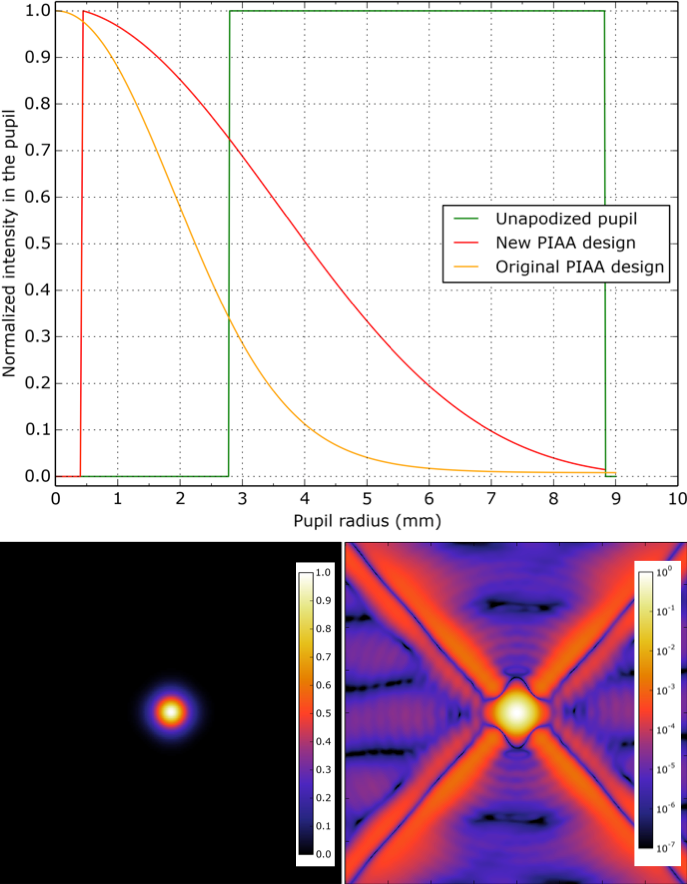}
\caption{\footnotesize (Top) Cross-sectional line profile of the intensity distribution of the unapodised pupil, pupil apodised with the original optics in SCExAO and the new apodisation optics. All profiles have been normalised to their peak. (Bottom) A simulated apodised PSF with the original PIAA optics (Left) linear stretch, and (Right) logarithmic stretch.}
\label{fig:apo}
\end{figure}

Pupil apodisation allows for the edges of the pupil to be softened, which eliminates the Airy rings in the focal plane around the PSF. A promising method that offers lossless pupil apodisation is via Phase Induced Amplitude Apodisation (PIAA) optics~\citep{guyon2003}. This was originally intended for high-contrast imaging of exoplanets at very small angular separation. The concept relies on geometrically remapping of the rays in the pupil in order to redistribute the light. This is typically done with two optics: An optic to push some of the rays at the edge of the pupil inwards and an optic to recollimate the beam. This was first practically demonstrated in the laboratory by~\cite{lozi2009} within an early version of the SCExAO testbed. The apodised pupil profile after the PIAA lenses utilised in SCExAO is shown in the top panel of Fig.~\ref{fig:apo} designated as the original PIAA design.  
It can be seen that the secondary obstruction is completely eliminated by these lenses as the profile has a peak at the centre of the pupil ($0$~mm). Also, the intensity of the beam decays at larger pupil radii but does not reach $0$. This was a deliberate feature of the design process. It was chosen to simplify the sag profiles of the two CaF$_{2}$ lenses to minimise features with small radii of curvature which were hard to manufacture by diamond turning at the time. Since the profile does not reach $0$, the system is designed to work in conjunction with a binary mask, a glass disk with thin concentric rings of opaque material which increase in number towards the edge of the pupil, that selectively attenuate the light (this is described in detail in~\cite{lozi2009}). In this way the apodisation profile is controlled in two stages. The binary mask by its very nature is a lossy device and reduces the $\sim86\%$ throughput of the uncoated PIAA lens pair to $\sim55\%$~\citep{jov2015} in H-band. The apodisation profile for a second generation PIAA design is also shown in Fig.~\ref{fig:apo} and designated as the new PIAA design. This design does not entirely eliminate the secondary but apodises the pupil in a single step, mitigating the need for the binary mask and hence preserving high throughput. This set of PIAA lenses has not been fabricated yet but with AR-coatings the throughput is expected to be $96\%$. The rest of this work focuses on the original PIAA lenses. Regardless, this shows that careful consideration of the total throughput of the system (throughput of optics and coupling efficiency) must be taken into account when designing an instrument (refer to section~\ref{sec:sumcoupopt} for details). 

An image of the simulated apodised PSF was calculated from the pupil illumination presented in the top panel of Fig.~\ref{fig:apo} and is shown in the bottom panel on both a linear (left) and log (right) scale. It is clear to see that indeed there are no diffraction rings about the PSF as expected and that the PSF is Gaussian-like (i.e. slowly decays away as one moves further off axis). It can also be seen that the diffraction due to the telescope spiders is unaffected by the apodisation process, as expected. 

\begin{figure}[!t]
\centering 
\includegraphics[width=0.99\linewidth]{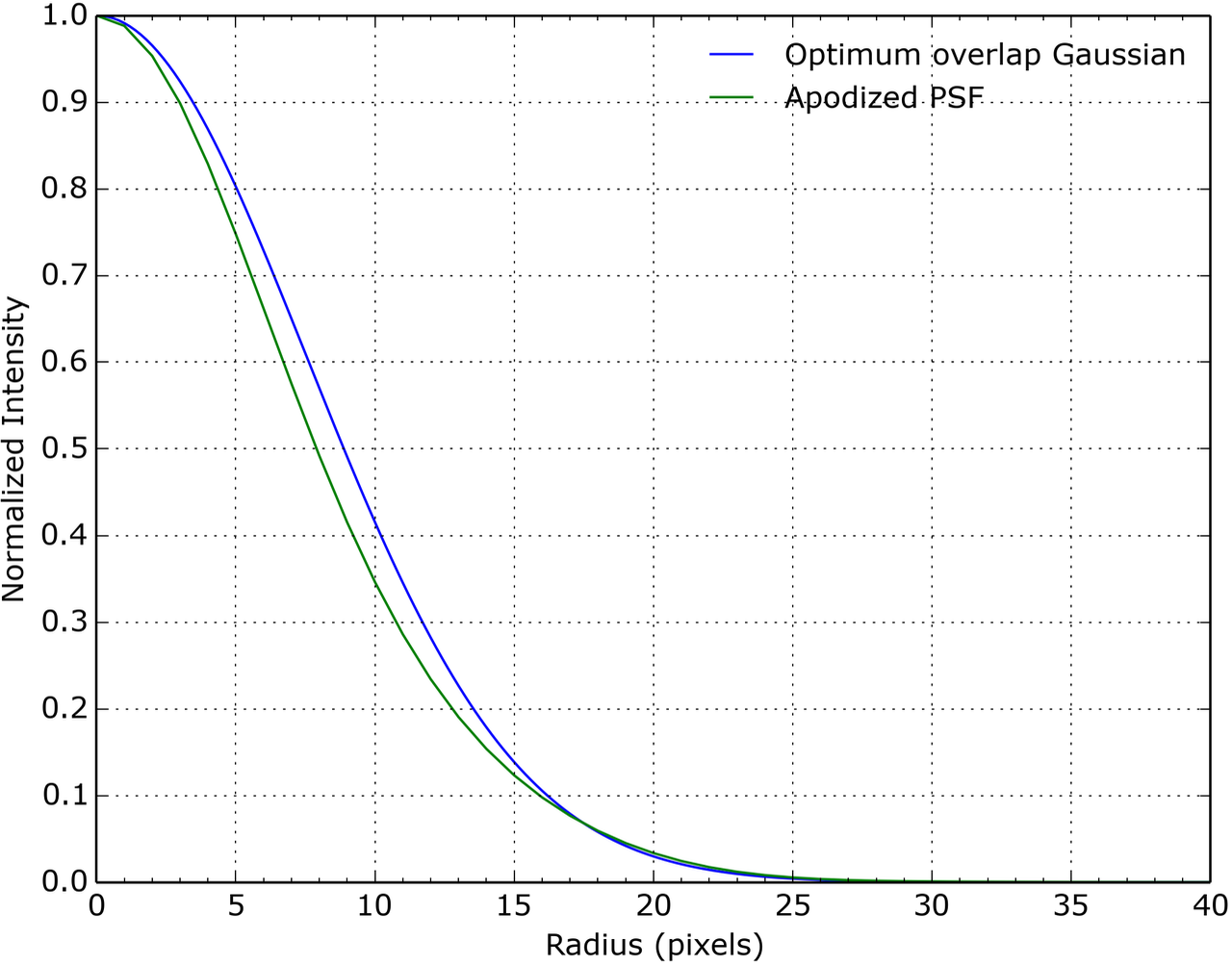}
\caption{\footnotesize Cross-sectional line profile of an apodised telescope beam and the optimum fitting Gaussian.}
\label{fig:apodized}
\end{figure}

The cross-sectional line profile of the apodised PSF and the optimum overlap Gaussian (which represents the mode of the fibre) are represented by the green and blue curves in Fig.~\ref{fig:apodized}, respectively. The apodised beam is a much better fit to the Gaussian mode of the fibre and indeed yields a vastly improved coupling efficiency of $\sim99\%$, a dramatic improvement over the $65\%$ for the Subaru Telescope pupil without apodisation. The coupling is systematically reduced by $\sim8\%$ once again when the telescope spiders are included in the simulation, confirming that indeed the apodisation had no effect on those. It is important to note that the PIAA optics must be redesigned for telescopes with different central obstruction sizes.

\subsection{Summary of coupling optimisation.}\label{sec:sumcoupopt}
Table~\ref{tab:throughputs} shows a summary of the results from the simulations described in~Section~\ref{sec:how}. These results are presented for the specific case of the Subaru Telescope central obstruction (size of $31\%$) including spiders and no wavefront aberrations, as this will be used throughout the remainder of this work. The table shows the theoretical coupling of the apodised and unapodised pupils into a SMF, reiterating the fact that apodisation is beneficial for boosting the coupling. The throughput of the pupil apodising optics is shown in the central portion of the table. A $100\%$ throughput is used to denote that there are no apodisation optics used in the unapodised case. It should be made clear that all other optics in the train are not considered in these values. The final portion of the table shows the combined throughput, defined as the multiplication of the coupling by the throughput of the apodising optics. It can be seen that at present, the large losses due to the binary mask in the original PIAA design outweigh the benefit of pupil apodisation at present. However, the new PIAA design is clearly more efficient and will significantly boost the overall throughput of the system in the near future. Despite the fact there is currently no advantage to using the original PIAA design, we choose to demonstrate the entire optical train as we envision it with the current optics to demonstrate the feasibility of the method.  

\begin{table}[ht!]
\centering
\caption{Coupling with and throughput of apodisation optics.}
\begin{tabular}{l c}
\hline\hline                                                            
(a) Coupling efficiency                         & $(\%)$                \\  [0.5ex]  \hline
Unapodised pupil                                        & $59$                  \\
\textbf{Original PIAA design}           & \textbf{$91$} \\
New PIAA design                                         & $91$                  \\      [1.0ex] \hline   
(b) Throughput of apodising optics      &                               \\  [0.5ex] \hline
Unapodised pupil                                        & $100$                 \\
\textbf{Original PIAA design}           & \textbf{$55$} \\
New PIAA design                                         & $96$                  \\      [1.0ex] \hline   
(c) Combined throughput (a$\times$b)&                           \\  [0.5ex] \hline
Unapodised pupil                                        & $59$                  \\
\textbf{Original PIAA design}           & \textbf{$51$} \\
New PIAA design                                         & $87$                  \\      [1.0ex] \hline
\end{tabular}\label{tab:throughputs}

\tablefoot{The table summarises the results from the simulations in Section~\ref{sec:how}. The theoretical coupling efficiencies are shown for three central obstruction sizes with no apodisation, the old PIAA lens design and the new PIAA lens design, for the case of no wavefront aberrations. The throughput of the PIAA lenses is presented as well as the combined throughput of each option, which is based on the throughput of the lenses multiplied by the coupling efficiency for each case.}
\end{table}

These simulations indicate that pupil apodisation optics optimised in consideration of the central obstruction of the telescope can greatly improve the theoretical coupling efficiency. The following sections outline the results of experimentally coupling light into SMFs with the original PIAA optics simulated here, and indicated by bold face text in the table.

\section{Experiment}\label{laboratory}
\subsection{The SCExAO instrument}
\begin{figure}
\centering 
\includegraphics[width=0.99\linewidth]{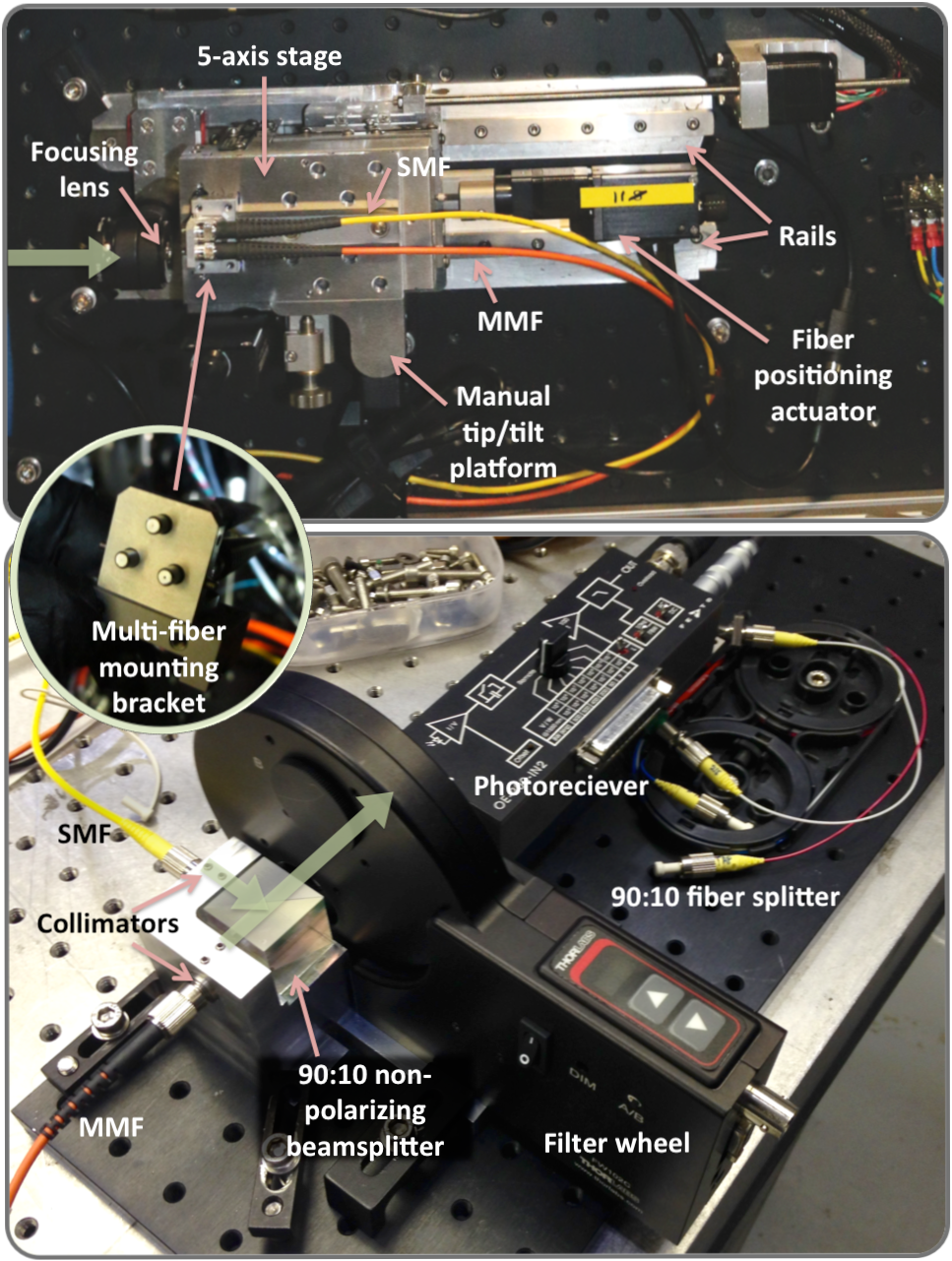}
\caption{\footnotesize (Top) Image of the fibre-injection mechanics. (Bottom) Image of the photometric setup. Transparent green arrows indicate the direction of the light path.}
\label{fig:rigs}
\end{figure}

\begin{figure*}[!ht]
\centering 
\includegraphics[width=0.85\linewidth]{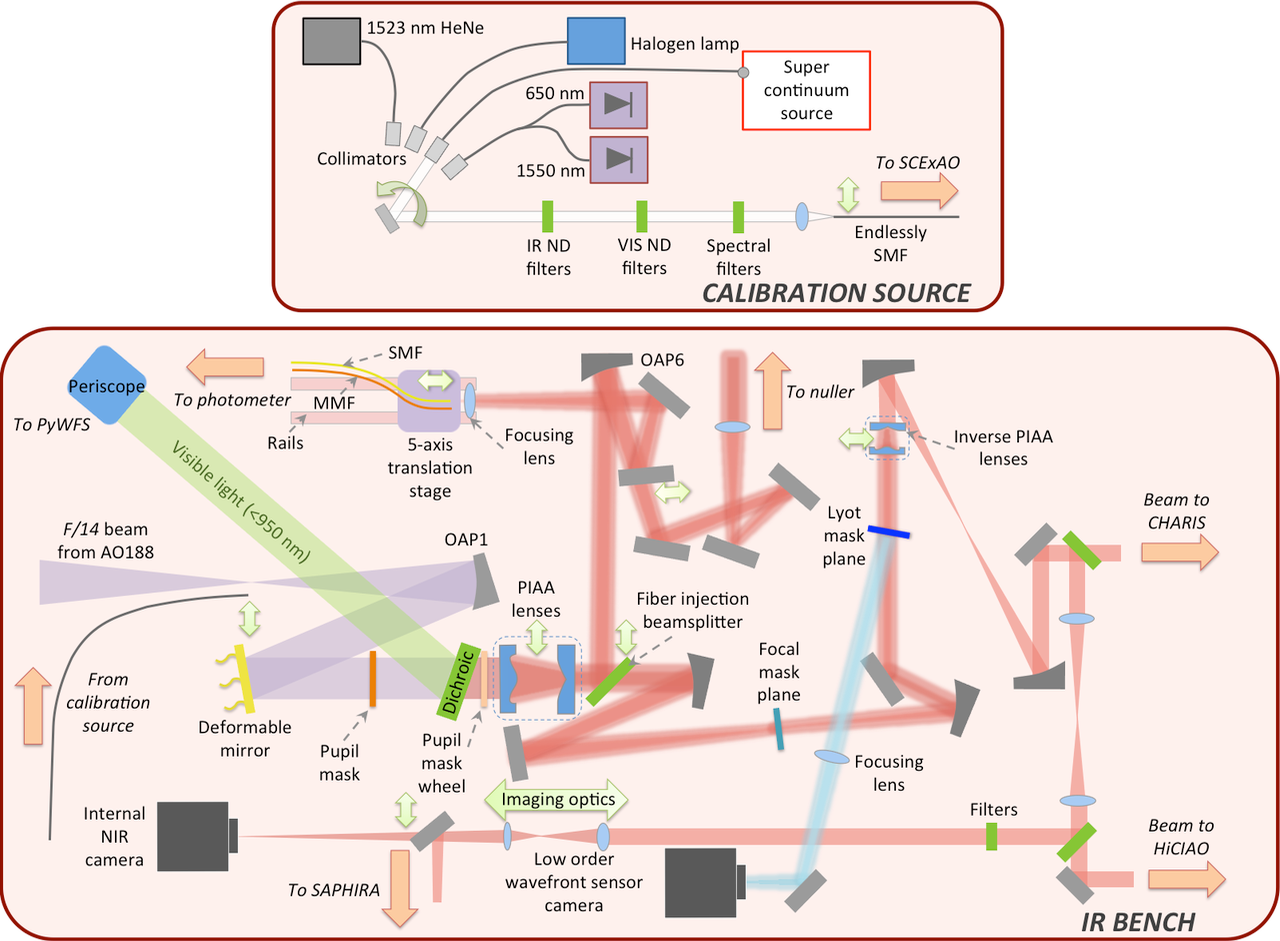}
\caption{\footnotesize Schematic diagram of the IR bench of the SCExAO instrument. (Top) Portable calibration source layout. (Bottom) IR bench layout. Dual head green arrows indicate that a given optic can be translated in/out of or along the beam. Orange arrows indicate light entering or leaving the designated bench at that location.}
\label{fig:schematic}
\end{figure*}

In this section we outline the experimental setup upon which the fibre injection concepts discussed in section~\ref{sec:how} were practically tested. An image of some of the hardware for the injection rig and the photometer is displayed in Fig.~\ref{fig:rigs}. A schematic layout of the testbed used, namely the SCExAO instrument, is shown in Fig.~\ref{fig:schematic}.

The SCExAO instrument was described in detail by~\cite{jov2015}; here we only focus on the features that pertain to this body of work. The light entered from AO$188$ (the facility AO system for Subaru Telescope) which offered a Strehl of $30-40\%$ in median seeing conditions in H-band~\citep{Minowa2010}. The light was first collimated by a gold-coated off-axis parabola (OAP) before the pupil was projected onto the $2000$ element, MEMS-based deformable mirror (DM) (Boston Micromachines Corporation). The light then passed through a fixed pupil mask which resembled the spider geometry of the telescope (with slightly oversized spiders for ease of alignment). Wavelengths shorter than $950$~nm were reflected by a dichroic immediately after the mask and directed up a periscope onto another bench to the pyramid wavefront sensor (PyWFS). The PyWFS is the high-order wavefront sensor which enables both higher spatial and temporal modes of the turbulence to be corrected than with AO$188$ alone. The PyWFS, still undergoing commissioning during the period of this body of work has achieved a Stehl ratio of up to $80\%$ in H-band and operates on $1000+$ modes at $2$~kHz with a latency of $\sim1$~ms. When fully commissioned the sensor will deliver a PSF with a $90\%$ Strehl ratio in the H-band in median seeing conditions on targets brighter than $9^{th}$ magnitude.

The infrared (IR) light transmitted by the dichroic was next incident on the PIAA lenses discussed in the section above. These can be deployed into the beam at any time or retracted entirely. An achromatic $90:10$ beamsplitter plate was used to direct $90\%$ of the flux from y- to K-band towards the fibre injection. This can be removed from the beam if needed. The beam was then focused by an OAP ($f=519$~mm) and directed by several flat mirrors to the focus. The throughput from the input of SCExAO to the fibre-injection rig was $78\%$ across J- and H-band without the PIAA and $43\%$ with the uncoated PIAA optics and the lossy binary mask. However, with anti-reflection (AR)-coated PIAA lenses which were designed to not make use of the binary mask, most of the $78\%$ throughput of the bench could be recovered in future. From~\cite{jov2015}, the total throughput from the top of the sky to the plane of injection in H-band was $\sim48\%$ without and $\sim26\%$ with the current PIAA optics. 

The beam that was transmitted by the beamsplitter was then directed via a series of gold-coated OAPs and mirrors to the internal near-IR camera (Axiom Optics, OWL SW1.7HS). The internal near-IR camera was used to align the instrument in both the pupil and focal planes and was the camera that was run to flatten the wavefront using the asymmetric pupil Fourier wavefront sensor~\citep{Mar2016}. For optimum performance of this sensor in the context of this work, the detector should have been located in the same plane as the SMF to minimise the effect of non-common path aberrations. Unfortunately however, this was not possible in our setup. The optimum coupling was determined while the PSF was aligned on a so called {hotspot} on the camera. This hotspot was used as a reference and when the PSF was returned to this position, the coupling was maximised into the fibre. 

Although not used in these experiments, a low-order wavefront sensor (LOWFS) could be used in the future to stabilise the PSF in regards to the low-order modes especially tip/tilt. More information is provided in section~\ref{LOWFS}. 

\subsection{Fibre injection assembly}
The injection consists of a five-axis stage (Newport, M-$562$-XYZ and $562$F-TILT) that allows the fibres to be translated around the focal plane, in X, Y, and Z with the aid of computer-controlled stepper motors (Zaber, T-NA08A25), which offered a minimum step size of $50$~nm and an unidirectional repeatability of $<1~\mu$m. Tip/tilt can also be adjusted via manual micrometers and was used to align the core of the optical fibre with the optical axis. This was done by looking at the beam in the pupil when reverse injecting red light into the fibre and steering the axes until the Gaussian illumination was centered on the pupil. This was only done once when the rig was first installed. In front of the rig, a $f=7.5$~mm achromatic lens (Thorlabs, AC050-008-C) was used to alter the speed of the beam as it entered the fibre. The fibre positioning rig and the lens were co-mounted on a carriage that rides on two parallel rails. By moving the carriage along the optical axis, that is, changing the distance between the small lens and the OAP, it was possible to adjust the focal ratio of the injection. The approximate range of $f/\#$ achievable was from $3$ to $25$. Atop the rig, a bracket was mounted that could host three SMA connectorised fibres at any one time. A standard telecommunications-grade SMF was used for the experiments (Corning, SMF-28-J9) as well as a large core ($365~\mu$m, NA=$0.22$), step-index multimode fibre (MMF) (Throlabs - FG365LEC) for calibration purposes. Using the stage, it was possible to manoeuver each of the fibres into the focus of the beam in turn. To swap between fibres, the stage was simply translated laterally.

The SMF-28-J9 fibre supported a $10.4~\mu$m $1/e^{2}$ mode field diameter at $1550$~nm. From the simulations carried out in the top panel of Fig.~\ref{fig:apodized}, the size of the optimum PSF was determined with respect to the Gaussian mode of the fibre. This was used to calculate that the optimum focal ratio for coupling efficiently into SMF-28-J9 at $1550$~nm was $f/5.3\pm0.1$ (we note this assumed that the apodised beam behaved as a Gaussian beam). The focal ratio of the beam was governed by the distance between the focusing OAP and the lens on the rig. To set the position of the small lens and hence obtain the desired $f/\#$, the following equation was used 
\begin{equation}\label{eq:focal_ratio}
s_{lens}=f_{lens}\left ( 1-\frac{f_{oap}}{\text{f\#} \cdot d_{oap}}\right ),
\end{equation}
where $s_{lens}$ is the distance of the small lens from the original focus of the OAP, $f_{lens}$ is the focal length of the small lens ($7.5$~mm), $f_{oap}$ is the focal length of the OAP ($519$~mm) and $d_{oap}$ is the diameter of the beam at the OAP ($\sim 8$~mm). This equation is based on the thin lens equation and was used to roughly position the small lens with respect to the OAP initially.

\subsection{Photometer}
\begin{figure}[!t]
\centering 
\includegraphics[width=0.85\linewidth]{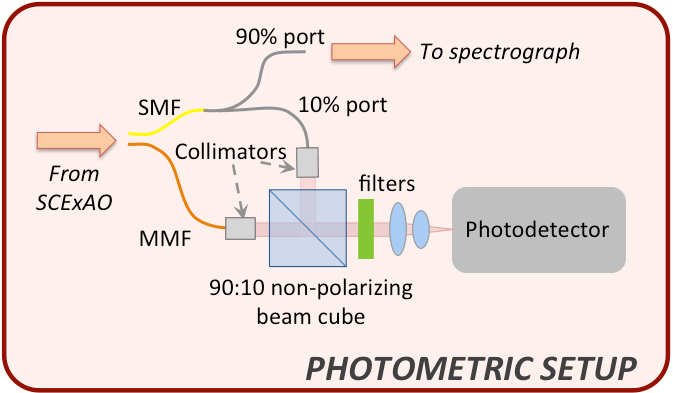}
\caption{\footnotesize The photometric setup.}
\label{fig:photometer}
\end{figure}

The SMF and MMF were routed to the photometer, which is shown in Fig.~\ref{fig:photometer}. The aim was to reimage both beams onto a single photodetector. The MMF was first collimated (Thorlabs, F220FC-1550) before $10\%$ of the light was transmitted through a $90:10$ beam cube and refocussed (using Thorlabs, AC080-020-C and C230TMD-C) onto a sensitive photodetector. The InGaAs variable-gain photoreceiver was a single-pixel device with a $300~\mu$m diameter and had a noise equivalent power of $7$~fW Hz$^{-1/2}$ in the highest gain setting (FEMTO, OE-200-IN2). 

The light in the SMF was first passed through an achromatic $90:10$ fibre splitter (Thorlabs, TW1550R2F2) where $10\%$ of the light was fed to the photometer. The beam was also collimated (Thorlabs, F220FC-1550) before $90\%$ was reflected by the beamsplitter cube directing it through the same focusing lenses onto the same detector. The dual lens focusing combination was used to ensure that the $365~\mu$m core size of the MMF was cleanly imaged within the footprint of the $300~\mu$m detector at all wavelengths. The $365~\mu$m core fibre was chosen based on the results of simulations presented in section~\ref{abscoup}. A set of narrowband filters ($25-50$~nm bandwidths) were used in an automated filter wheel spanning J- and H-band that ranged from $1250$ to $1600$~nm. The $90\%$ port of the splitter was routed towards a compact photonic spectrograph that is not discussed here (refer to~\cite{jov2016a} for further information). By using the two splitters in this way, it was possible to obtain similar levels for the flux between the SMF and MMF, which was a necessity when trying to keep both signals within the limited dynamic range of the detector.  

The MMF was used to calibrate the absolute flux in the SMF. This was done by first collecting a data set with the SMF and then moving the MMF into the focus, and dividing one signal by the other. This also meant only one fibre was in the focus at any one time (single source observations). 

Data sets of $15$~s in duration were collected for each spectral channel and fibre type, with a sampling of $1$~kHz. After a data set was collected, a background signal where no flux was landing on the detector was also collected. The background signal was averaged and then subtracted from the data before any further calculations were undertaken. The processed data was then used to determine the coupling efficiency presented in the following section.

\section{Laboratory results}\label{labresults}
\subsection{Coupling efficiency into a SMF}
With the aim of determining the practical limitations of coupling into a SMF from an apodised telescope pupil, a detailed laboratory characterisation was undertaken as outlined in this section. To ensure maximum coupling, the wavefront was first flattened using the asymmetric pupil Fourier wavefront sensor~\citep{Mar2016}. This was done in the H-band (from $1500$--$1650$~nm) using the near-IR internal camera as outlined above. Although this minimised the chromatic residuals, it did not necessarily address non-common path aberrations as the detector was not located in a common focal plane with the fibre injection. This was a good starting point as all optics in SCExAO were specified to a flatness better than $\lambda/20$ at $633$~nm (RMS) minimising non-common path errors by design (discussed further below).

\begin{figure}[b!]
\centering 
\includegraphics[width=0.99\linewidth]{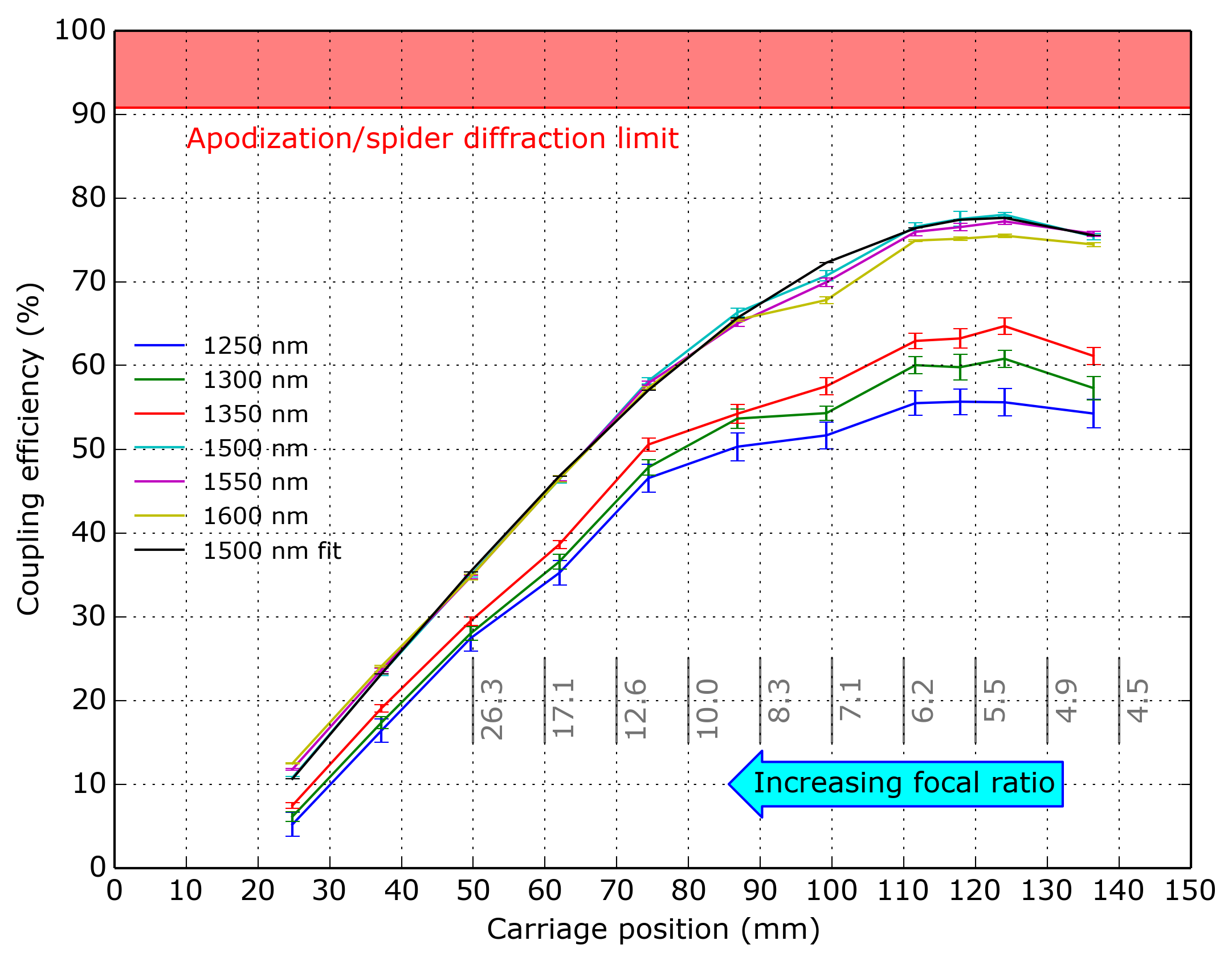}
\caption{\footnotesize Coupling efficiency as a function of the fibre injection rig carriage position, which is related to the focal ratio. The estimated focal ratio for each position of the rig is indicated next to the vertical grid lines. This was calculated assuming that peak coupling occurred at $f/5.3$.}
\label{fig:coupvsfr}
\end{figure}

The light coupled into the SMF was optimised in X, Y, and Z using the photometer. The carriage supporting the injection was translated along the rails varying the $f/\#$. At each point, the flux was optimised for the SMF and MMF and data sets (described above) were collected in all spectral channels. The coupling efficiency as a function of the focal ratio is shown in Fig.~\ref{fig:coupvsfr}. 
We note that the loss due to Fresnel reflection from the input and output surface of the fibre ($\sim4\%$) has been removed from these values. The error bars are based on the fluctuation in the power during data acquisition. This is primarily comprised of the relative motion of the injected beam with respect to the SMF core due to vibrations at the Nasmyth platform and/or gusts from the air conditioning creating bench turbulence, but also consists of minor fluctuations in the light source as well. 

The figure highlights several key features. Firstly, the coupling efficiency was optimum at a narrow range of carriage positions (i.e. focal ratios) and degrades on either side of this. Although not directly validated, the peak in the curve for $1550$~nm at $\sim124$~mm is likely to correspond to the optimum focal ratio of $f/5.3$ or thereabouts. Secondly, the coupling efficiency was not equal at all wavelengths. Indeed the highest coupling efficiency of $\sim78\%$ was at $1500$/$1550$~nm and reduced to $55\%$ at the shortest wavelengths. This chromatic coupling could be the result of a number of things. The first is a differential dispersion in the mode field diameter size and the beam size in the focal plane. The second is residual wavefront error, which is larger at shorter wavelengths. However, this could not account for the full $30\%$ difference in coupling between H- and J-bands. Finally, it could be due to chromatic aberrations in the PIAA lenses. Although the lenses were fabricated from CaF$_{2}$, there may be some uncharacterised chromatic aberration. Regardless, the coupling efficiency was high across all wavelengths. This leads to the third point; the coupling efficiency almost reaches the theoretical limit of $\sim91\%$ (highlighted by the shaded-in red region in Fig.~\ref{fig:coupvsfr}) taken from Table~\ref{tab:throughputs}. Indeed, the data shows that a coupling of $86\%$ of the theoretical limit was achieved. This minor loss in coupling could be attributed to some non-common path errors between the asymmetric pupil Fourier wavefront sensor detector plane and that of the fibre injection and/or imperfections in the PIAA optics. Static Zernike mode profiles were applied with varying amplitudes to the DM in linear combinations by a downhill simplex algorithm and it was determined that there were negligible non-common path errors to first order. A more careful scan with higher-order modes should be conducted. This may improve the coupling, but here we place an upper limit on the level of improvement that would be possible of a few percent at best. Imperfections in the PIAA optics are believed to be the dominant source of coupling efficiency degradation. However, it is difficult to determine if there are imperfections from the PSF image alone and use this to determine where it comes from and quantify the impact so we leave this as an item for a future exploration. 

Figure~\ref{fig:coupvsfr} shows the results in the regime with a near perfectly flat wavefront. However, for use on-sky it is important to understand the performance of the injection as a function of the wavefront residuals after AO correction. This was tested in a laboratory setting by using SCExAO's turbulence simulator. An overview is presented in section~\ref{turbsim} while full details of this can be found in~\cite{jov2015}. Data was collected with the photometer while the turbulence simulator was adjusted to operate over a range of settings corresponding to various levels of wavefront correction. 

\begin{figure}[!b]
\centering 
\includegraphics[width=0.99\linewidth]{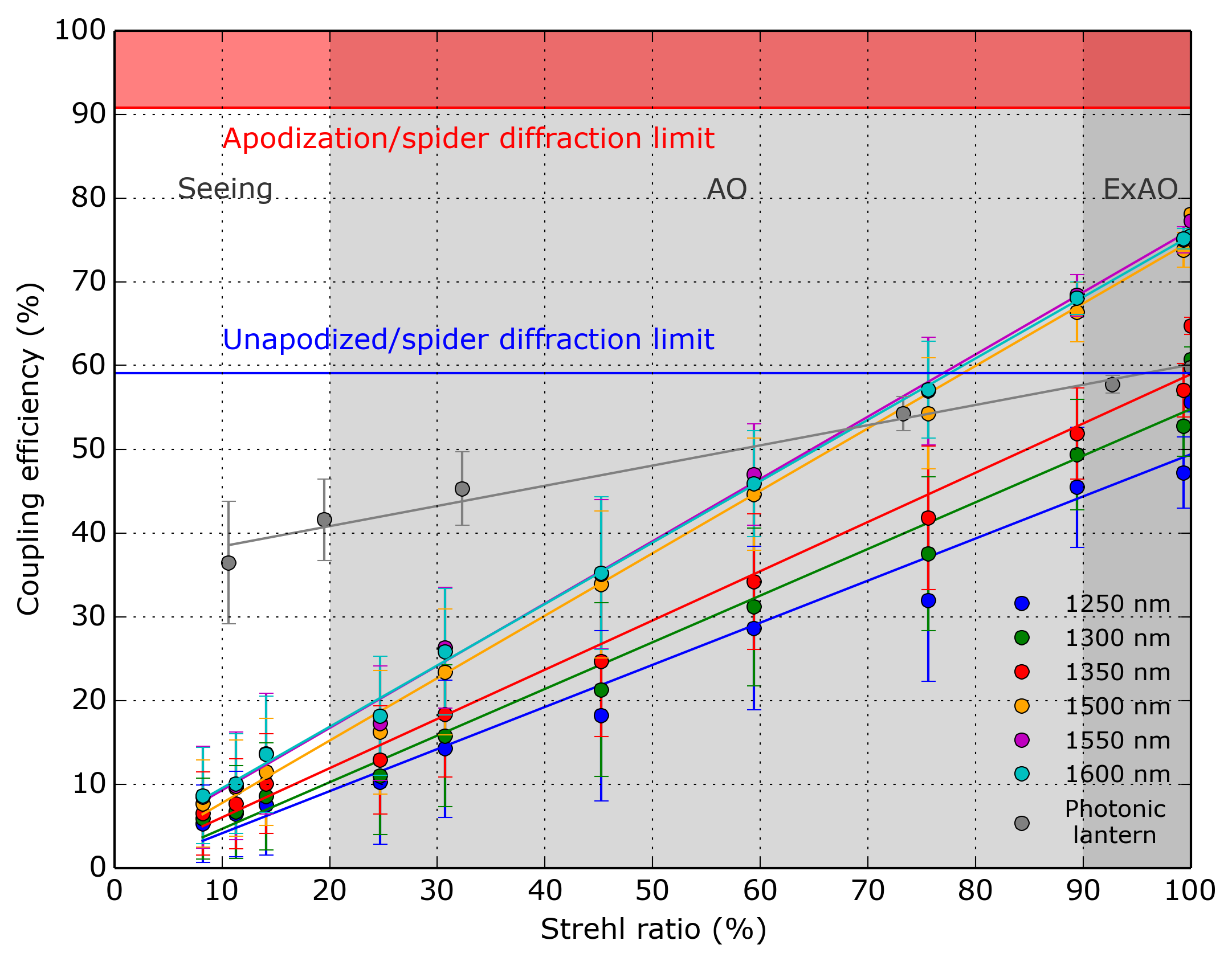}
\caption{\footnotesize Coupling efficiency as a function of the measured Strehl ratio in H-band ($1500$-$1650$~nm). The ExAO, AO and seeing regimes are highlighted for convenience. The coupling efficiency limit due to spiders and apodisation optics is highlighted in red, while the limit without apodisation is in blue.}
\label{fig:coupvssr}
\end{figure}

Figure~\ref{fig:coupvssr} shows the coupling efficiency as a function of the measured Strehl ratio. The same result plotted as a function of the calculated RMS wavefront error is presented in Fig.~\ref{fig:coupvsrms} in section~\ref{coupvrms} for completeness. 
It can be seen that the coupling efficiency ($\eta$) increases linearly as a function of Strehl ratio, which is an important relationship to understand for future instrument development. The equation of the line of best fit for the data at $1550$~nm is given by $\eta = Strehl~Ratio \times0.74+1.84$ ($\%$). We note the Strehl ratio needs to be input as a percentage and $\eta$ will be expressed as one. Also, the error bars increase as the Strehl ratio decreases. This is because at lower Strehl, the instantaneous Strehl fluctuates as a result of the varying wavefront error, reducing the stability of the coupling. The theoretical maximum coupling efficiency for the Subaru Telescope pupil geometry assuming no apodisation was implemented is shown in both figures (blue horizontal line) for reference. By using the apodisation optics, we have surpassed the maximum coupling possible without the lenses, demonstrating this application of the lenses for the first time and justifying their use. Indeed, the performance of a perfectly optimised injection assuming the lenses were not used would be surpassed by a system using the lenses and a beam degraded to a Strehl ratio of $75\%$ (between $1500$ and $1600$~nm). With ExAO levels of wavefront correction ($90\%$ Strehl ratio) a coupling efficiency as high as $67\pm2\%$ could be achieved in the range of $1500$--$1600$~nm. This is a highly efficient injection efficiency that could be justified in future instrumentation if access to an ExAO system were possible.        

Another useful value to keep in mind for several reasons is the coupling efficiency of $50\%,$ which can be achieved with a Strehl ratio of $65\%$. Firstly, this value is within a factor of $1.8$ of the theoretical maximum ($91\%$) which is achieved by significantly improving the wavefront error and pushing the Strehl towards $\sim90\%$ on-sky (as well as addressing the unknown losses). To do this however, requires complex multi-layered wavefront control systems and median, or better than median seeing, free from telescope tracking limitations/vibrations. Sacrificing a factor of $\sim2$ in flux in order to minimise the complexity of the AO loop control and/or access nights without better than median seeing (i.e. more nights) would be an acceptable trade for most targets. Secondly, this value is within a factor of $2$ of the maximum achievable for the MMF used in this experiment (much larger than those typically used for spectroscopy), and of the order of the same level of coupling typically obtained on-sky for MMF-fed spectrographs (this depends on how the PSF, AO-corrected or not, is matched to the core size of the MMF and varies on a case-by-case basis). Taking this into account it is clear that it would be beneficial to use a SMF instead of an MMF to feed a spectrograph and eliminate modal noise if operating in even a $50\%$ Strehl-ratio regime, which can be achieved by conventional AO systems in good seeing conditions or at longer wavelengths.

\subsection{Coupling efficiency into a mode-converting photonic device}
A photonic lantern is a mode-converting device that splits the light from a single MMF to multiple SMF outputs~\citep{saval2013}. The device was embraced by astronomers to convert the seeing-limited light of a telescope to multiple diffraction-limited beams so photonic components could be exploited~\citep{jbh2011}. In the context of this work, a photonic lantern would relax the requirements on mode matching as the input is a MMF but deliver diffraction limited performance. It is, therefore, interesting to see how it compares. 

\begin{figure}[!h]
\centering 
\includegraphics[width=0.85\linewidth]{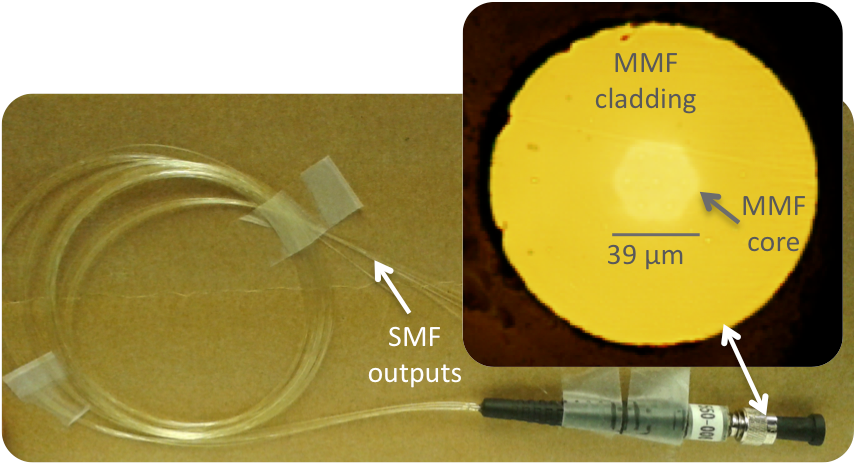}
\caption{\footnotesize Image of the $1$ by $7$ photonic lantern. (Inset) A microscope image of the MMF end of the photonic lantern.}
\label{fig:pl}
\end{figure}

To examine this, the coupling efficiency was measured for a device that consisted of seven SMFs, typically referred to as a $1$ by $7$ photonic lantern. The device was constructed by tapering down seven, SMF-$28$ fibres in a single fluorine-doped capillary tube. The core size of the MMF end was $\sim39~\mu$m, and based on this and the known refractive index contrast for the core, simulations predicted that the lantern supports the right number of modes to facilitate an efficient transition to the SMFs. An image of the photonic lantern used and a microscope image of the MMF end of the lantern are shown in Fig.~\ref{fig:pl}. 

The SMFs at the output of the lantern were spliced to a V-groove array (an array which hosts several fibres in a linear array where the output facet of all fibres is terminated in a flat polish). The coupling efficiency for the lantern at $1550$~nm was measured using a power meter instead of the photometer setup and the light was injected with the optimum focal ratio for the SMF (i.e. $\sim5.3$). The results are shown in Figs.~\ref{fig:coupvssr} and \ref{fig:coupvsrms} in grey. It is clear that the coupling efficiency has a near-linear dependence on Strehl ratio as was the case for the SMF, albeit with a shallower slope. Although the peak coupling efficiency in the diffraction limit is only $60\%$ as compared to the $80\%$ for the SMF, the coupling drops off slower as a function of reduced Strehl, and so even at very low Strehl ($10\%$ for example), a coupling efficiency of nearly $38\%$ can be maintained. This is attributed to the fact that the lantern can support more than one mode enabling more efficient capture of the light when it is not well confined to the core of the PSF. It is important to clarify that the photonic lantern used in these experiments was not optimised at the fabrication stage for this work. Also, the focal ratio was not optimised for the experiments, and one would expect that a higher focal ratio would be more optimal to match to the fundamental mode of the lantern based on mode profile calculations for the tested device. In fact, the simulations revealed that the fundamental mode is not perfectly circularly symmetric, which could place an upper limit on the maximum coupling possible to the apodised beam, but the loss associated with this is not expected to be more than a few percent. Finally, the effect of optimising the wavefront at the entrance to the fibre with the DM was also not investigated. Therefore the peak coupling of $60\%$ is a lower limit to what can be achieved.

\section{On-sky results}\label{sec:onskyresults}
The performance of the fibre injection was tested on-sky in March $2016$ during the SCExAO engineering run. This was done to 1) validate the performance of the injection system in a real world setting and 2) to confirm that the relationship between Strehl ratio and coupling holds and can be exploited for planning observations. The PyWFS on SCExAO was undergoing commissioning at the time and was not fully operational, but did offer an improvement on the Strehl ratio over AO$188$ alone. It has achieved a Strehl ratio as high as $80\%$ in the H-band but cannot currently support this level of performance in anything but better than median seeing conditions during periods free from vibrations. 

A preliminary data set collected during periods of strong telescope vibration on the $18^{th}$ of March can be found in section~\ref{apponsky}. On the night of the $21^{st}$ of March, $2016$, the conditions were photometric and the seeing was $0.6"$ in V-band. The fibre injection was tested on Alpha Hydrae (K$3$ spectral type, R-mag$=0.93$, H-mag$=-1.05$). Before inserting the injection, an image of the PSF was recorded for $2$ minutes to monitor wavefront control performance. The average Strehl ratio during this period was $69.0\pm3.4\%$ (in H-band). After the fibre injection tests were complete, a single Strehl ratio measurement was made that indicated that the Strehl had dropped to $57.5\%$. The conditions may have deteriorated slightly during the period of data collection and therefore we take the average of these values, $63\pm7\%$ to be the approximate Strehl ratio during the time of data collection. The star was steered onto the \textit{hotspot} that was pre-registered on the internal NIR camera. The PIAA optics and mirrors were inserted to divert the light to the fibre injection. Flux was detected on the photometer. The fibre position was tweaked slightly to maximise flux. No active tracking of the fibre position was implemented. 

\begin{figure}[!ht]
\centering 
\includegraphics[width=0.99\linewidth]{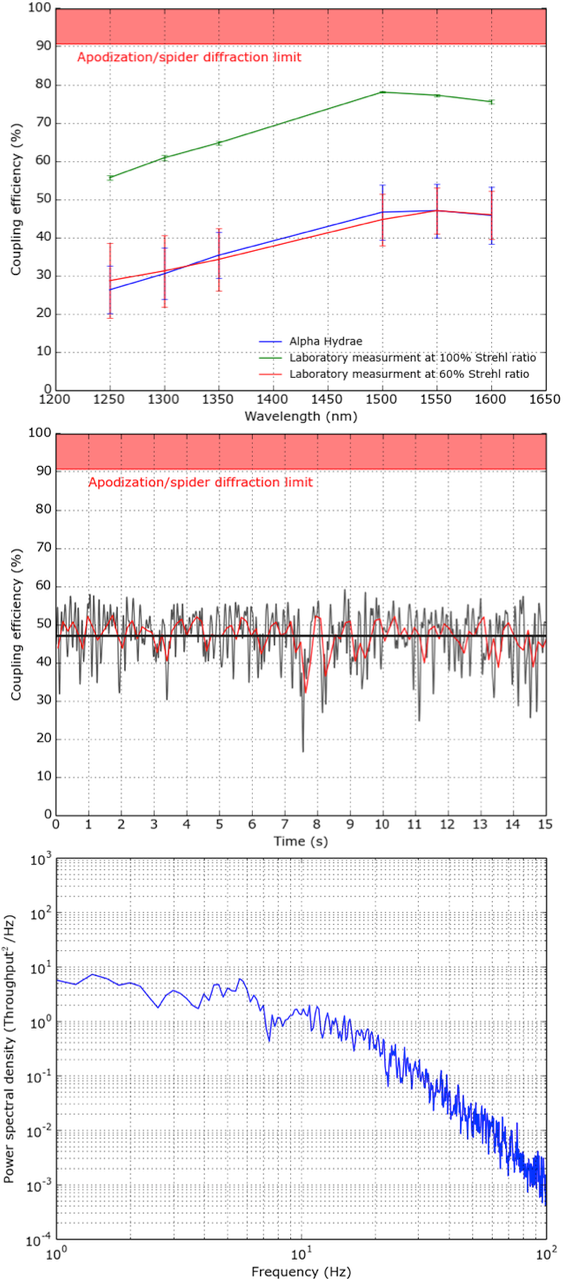}
\caption{\footnotesize (Top) Coupling as a function of wavelength (target: Alpha Hydrae taken on the night of the 21st of March). The green line shows the best laboratory result obtained with no wavefront error applied. (Middle) A time series of the 1550 nm spectral channel. The dark solid line shows the mean over the 15 s data-collection period. The red line shows the data binned into 150 data point bins. (Bottom) A PSD of the time series in the middle panel. }
\label{fig:onskyalphhya}
\end{figure}

Numerous data sets were collected on this target and the best overall coupling over the entire spectrum at one instance in time is presented in blue in the upper panel of Fig.~\ref{fig:onskyalphhya}. 
\begin{figure}
\centering 
\includegraphics[width=0.99\linewidth]{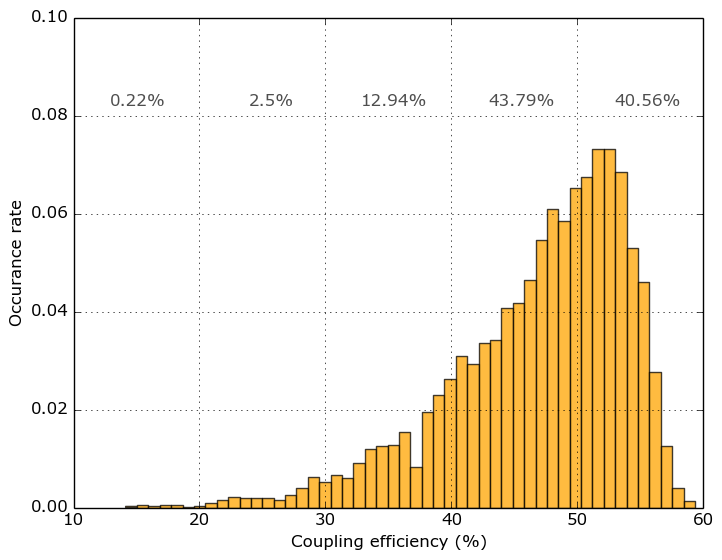}
\caption{\footnotesize The occurrence rate as a function of coupling efficiency for the data collected on Alpha Hydrae. The cumulative total occurrence for each $10\%$ coupling efficiency band starting from $10\%$ is shown in grey (i.e. bin one is $10$--$20\%$, bin 2 is $20$--$30\%$ and so on).}
\label{fig:onskyoccurate}
\end{figure}
The average coupling was $\sim47\%$ between $1500$ and $1600$~nm and tapered off slowly at shorter wavelengths as previously seen in the laboratory. However, a decent level of coupling was maintained down to the shortest wavelength. From Fig.~\ref{fig:coupvssr} a coupling efficiency of $47\%$ at $1550$~nm corresponds to an average Strehl ratio of $60\%$ in H-band, which is consistent with the value computed from the focal plane images around the time of data collection. The laboratory data for a Strehl ratio of $60\%$, presented in Fig.~\ref{fig:coupvssr}, are overlaid on Fig.~\ref{fig:onskyalphhya} with a red line. A good agreement can be seen between the laboratory and on-sky data sets given the measured Strehl ratio of $60\%$, verifying that the relationship portrayed in Fig.~\ref{fig:coupvssr} holds on-sky. 

The middle panel in Fig.~\ref{fig:onskyalphhya} shows a typical time series in the $1550$~nm spectral channel. Overlaid on the data is a solid line which represents the average coupling over the $15$~s data collection window. Also, the red line shows the data with bins consisting of $150$ data points each. The time series reveals that the coupling mostly varies between about $40$ and $55\%$ and undergoes rapid short-lived drops below this level (which corresponds to a Strehl ratio range of $52$ to $74\%$). This can be seen clearly in Fig.~\ref{fig:onskyoccurate} which shows a histogram of the time series data. Indeed, the coupling efficiency was above $50\%$ almost $41\%$ of the time and above $40\%$ for $84\%$ of the time. At some point it even approached $60\%$ (Strehl ratio of $80\%$). A PSD of the time series is shown in the bottom panel of Fig.~\ref{fig:onskyalphhya} and reveals no specific resonance during this data set.

\section{Discussion}\label{sec:discussion}
The on-sky results are consistent with the relationship shown in Fig.~\ref{fig:coupvssr}. This is important as it means that one can use knowledge of the Strehl ratio to determine what level of coupling could be obtained from the fibre injection on any given night. The relationship in Fig.~\ref{fig:coupvssr} only applies to the fibre injection in SCExAO and its systematics and a similar relationship should be developed for any other AO system/fibre injection of interest. Once this relationship has been developed, it is possible to use it deterministically to calculate expected levels of coupling, which can be used in future instrument design concepts or even for data reduction.  

In addition, the results presented in Section~\ref{sec:onskyresults} highlight that the coupling efficiency on-sky is completely limited by Strehl. With higher levels of Strehl ratio available, higher coupling efficiencies can be achieved. These results also compare well with previous on-sky attempts at coupling light into SMFs with non-apodised beams. 

If the stability of the PSF in this work were improved by, for example, improving the sensitivity of the PyWFS to tip/tilt, deploying a Linear Quadratic Gaussian (LQG) control loop, which is designed to use predictive control to drive the PyWFS to notch out any resonances~\citep{poyneer2014}, and/or running the LOWFS, the coupling would be more stable with time. This would also make it easier to tweak and optimise the alignment of the fibre on-sky in order to maximise the flux. These are future upgrades that will no doubt improve performance, but we have shown with preliminary on-sky data that even with a non-fully operational extreme AO system, efficient coupling into a SMF from a large telescope ($8$-m class) and apodisation optics is possible and efficient. 

The data presented in Fig.~\ref{fig:coupvssr} allow a future instrument designer to consider the relative merits of a SMF versus a photonic lantern for their instrument feed for spectroscopy. The photonic lantern subdivides the flux on average by the number of SMF ports. This means that although the coupling is relatively high ($38\%$) into the lantern at a $10\%$ Strehl ratio, the signal in each port is a factor of $7$ lower (equivalent to a coupling of $\sim5\%$ into each of the SMF ports) when the light is evenly distributed amongst the cores. This in fact is lower than the coupling directly into a SMF with the same Strehl ratio ($10\%$) and hence the S/N for a single SMF would be higher than that for each of the ports of the lantern. This raises the first important point: One must carefully consider the dominant noise source in their observations. If the star is bright and photon noise dominates, then the seven ports of the lantern could be co-added together in post-processing and the S/N would be improved by a factor of seven. On the other hand, if read noise is the dominant noise source, then co-adding the signal in the ports would improve the S/N by $\sqrt{N}$ or $2.6$ in this case. Of course one must consider the dark noise as well and ensure that the S/N in each lantern port is sufficient to be above the dark noise. The second important point is that a lantern, by default, requires more pixels than a SMF for an equivalent instrument specification (i.e. spectral sampling). The example of even illumination amongst the seven cores used above is a little unfair in that the coupling to the fundamental mode of the lantern will be greater for higher Strehl ratios (i.e. when there is a PSF core it will have a higher overlap with the fundamental mode than to the others). The fundamental mode of the multimode end of the lantern has a near perfect correspondence to a single SMF core at the output based on symmetry and an optimised device design. This means that one port will have a much better S/N than the others and could potentially be on par with a SMF. This should be taken into consideration as well when choosing the optimum collection device for an instrument. This illustrates that although photonic lanterns are the only efficient counter option to a SMF for getting the light into a diffraction-limited mode at Strehl ratios $<10\%$, it is important to consider the dominant noise process and pixel availability. However, new photonic lantern developments in the area of few-mode optical communications may offer a more optimised device design. For instance, mode-selective photonic lanterns~\citep{saval2014} could enhance coupling conditions into a smaller number of SMFs ($1$-$3$ compared to $7$ in the current study) while maintaining the demonstrated higher coupling efficiencies at lower Strehl ratios.

\section{Summary}\label{sec:summary}
In this body of work, we outline and demonstrate how the combination of lossless pupil apodisation optics and advanced wavefront control can be used to efficiently inject light into a single-mode device, be it a fibre or waveguide architecture, for the first time. The PIAA optics used for pupil apodisation increased the theoretical coupling efficiency of the telescope beam into a SMF from $59\%$ to $91\%$. In a laboratory setting, we achieved a coupling efficiency at $1550$~nm of $86\%$ of the theoretical limit. A study was conducted into the effect of Strehl ratio on coupling and it was determined that there was a linear correlation. This trend was reinforced by the on-sky data allowing for the expected coupling to be determined from the Strehl ratio which is a powerful tool for future instrument design. For ExAO levels of Strehl ratio ($90\%$ in H-band) the coupling efficiency should be above $67\%$, which is similar to a typical MMF injection efficiency for currently operating spectrographs. For high levels of AO correction (Strehl ratio $\sim50\%$ in H-band), possible in better-than-median seeing conditions on most AO systems (or at longer wavelengths), a coupling of $40\%$ can be achieved. For standard AO correction (Strehl ratio $20$--$40\%$), it should be possible to achieve $>18\%$ coupling, which is sufficient for brighter targets. Few-port photonic lanterns become the obvious choice over SMFs for Strehl ratios $<20\%$, but the S/N and pixel availability must be taken into consideration. These devices should not be overlooked for use in the high-Strehl-ratio regime as well. The preliminary on-sky SMF results are encouraging with a $15$~s time-averaged coupling efficiency of $47\%$ achieved around $1550$~nm. In this case the coupling was $>40\%$ at $1550$~nm $84\%$ of the time.   

The results clearly show that the coupling efficiency into a SMF is no longer a barrier to exploiting diffraction-limited devices for spectroscopy or interferometry. Future instrument designs should consider the possibility of employing a SMF feed to their instrument to take full advantage of the spatial filtering and temporally invariant PSF that this offers. Indeed, the next-generation high-resolution spectrograph for the LBT, iLocator, is doing just this~\citep{crepp16}. It will be the first instrument to be purpose built to use a SMF feed. A comprehensive overview of the advantages this presents in the context of stellar spectroscopy was recently given by~\cite{jov2016b}. In addition, the KPIC project at the Keck Telescope will go one step further and exploit the superior speckle rejection properties of a SMF to feed NIRSPEC with SMFs to enable the direct characterisation of known exoplanets~\citep{Wang2017,dimitri2017}. Finally, embracing a SMF-feed in future instrument design will also help to minimise the footprint of the instrument through the ability to exploit advanced photonic technologies.

\begin{acknowledgements}
The authors acknowledge support from the JSPS (Grant-in-Aid for Research \#$23340051$, \#$26220704$ \#$23103002$). This work was supported by the Astrobiology Center (ABC) of the National Institutes of Natural Sciences, Japan and the directors contingency fund at Subaru Telescope. This research was also supported by the Australian Research Council Centre of Excellence for Ultrahigh bandwidth Devices for Optical Systems (project number CE110001018). The authors wish to recognise and acknowledge the very significant cultural role and reverence that the summit of Maunakea has always had within the indigenous Hawaiian community. We are most fortunate to have the opportunity to conduct observations from this mountain. 
\end{acknowledgements}

\section{Appendix}\label{sec:appendix}

\subsection{Limit of coupling an Airy beam into a SMF}\label{couplimit}
To understand this limit, it is instructive to study the bottom panel of Fig.~\ref{fig:aberrations}, which shows the cross-sectional line profile of an Airy pattern (red curve) and the corresponding best-overlap Gaussian (yellow curve). The profile of the optimum Gaussian is nearly $0$ ($<1\%$) around the first null in the Airy pattern. This indicates that optimum coupling is achieved when only light from the PSF core is collected by the fibre. The reason behind this is that the light in the first ring is out of phase with that of the PSF core, as described. Hence, if it were to be injected into the Gaussian mode, it would lead to destructive interference and a reduction in the coupling efficiency (first described by~\cite{Shaklan1988} and~\cite{Coude1994}). Therefore, optimum coupling between an Airy pattern and a Gaussian occurs when the Gaussian is well matched to the size of the Airy core, but does not include light from the rings.

\subsection{Eliminating losses due to spiders}\label{spiders}
A method for eliminating the presence of the spiders was proposed by~\cite{lozi2009}. They outlined the concept of a spider removal plate which consists of four equally thick, flat pieces of glass that are used to shift the four illuminated regions of the pupil (between the spiders) inwards reducing the effective size of the spiders. In practice this is done by tilting each one of the parallel pieces of glass at a constant angle with respect to the optical axis in a collimated space. However, the device they demonstrated introduced aberrations to the beam so it was deemed unfit for use and has therefore not been used in the work presented here either. With improvements in the manufacturing process this could be used in future.

\subsection{Low-order wavefront sensing}\label{LOWFS}
 The LOWFS uses the light diffracted by a mask in the focal plane down stream of the beamsplitter to monitor the changes to the shape of the PSF~\citep{singh2014}. The diffracted light is reflected by a reflective Lyot stop towards the LOWFS camera, operating with light from Y- to H-band. This is primarily used for coronography and has achieved an on-sky tip/tilt residual of $0.15$~mas in H-band~\citep{singh2015}. The ability to stabilise the tip/tilt at the wavelength of operation, close to the focal plane of injection and at a speed of up to $170$~Hz is valuable for efficient long term coupling and will be explored in detail in the future.

\subsection{Absolute flux calibration}\label{abscoup}
\begin{figure}[!ht]
\centering 
\includegraphics[width=0.99\linewidth]{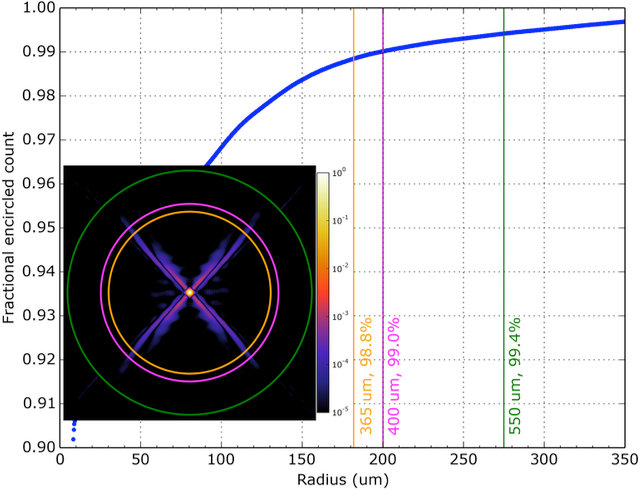}
\caption{\footnotesize The fraction of encircled counts of the apodised PSF as a function of off-axis distance in microns. (Inset) The apodised PSF logarithmically stretched with overlaid circles showing the relative size of various MMFs.}
\label{fig:coupinMMF}
\end{figure}

The MMF was used to calibrate the absolute flux in the SMF. This was done by first collecting a data set with the SMF and then moving the MMF into the focus, and dividing one signal by the other. This also means only one fibre was in the focus at any one time (single source observations). Simulations were carried out to see what size of fibre would be sufficiently large enough to capture the entire PSF, including the diffraction features of the spiders as well as light spread into the halo by, for example, bad seeing/bad AO correction. It was determined that $99\%$ of the flux could be collected by a $365~\mu$m core size fibre as shown in Fig.~\ref{fig:coupinMMF}, and using larger fibres would not improve this by much but would complicate the optics in the photometer setup. Therefore this size was chosen. The fibre had an approximate extent of $2.4"$ on-sky. This is approximate as the fibre is used in a post-apodised plane and there is a pupil size and shape change upon remapping that needs to be accounted for accurately. 

\subsection{Turbulence simulator and measuring the Strehl ratio}\label{turbsim}
The turbulence simulator uses the DM to generate a constantly moving phase screen, based on a Kolmogorov turbulence profile. The wind speed can be controlled by changing the speed of the phase map passing over the DM. The amplitude of the phase errors can be controlled by the amplitude applied to the DM. Also, the amount of low-order aberrations can be controlled with a single scaling coefficient, that is, a multiplicative factor applied to the lowest Fourier modes of the Kolmogorov turbulence profile. In this way it is possible to mimic the effect of an AO system that typically operates on the lower-spatial-frequency modes of the atmosphere.

To accurately determine how much wavefront error was applied in each case, non-PIAA images were collected on the internal NIR camera for each setting of the turbulence simulator. A cube of $1000$ images was first dark subtracted before collapsing into a single frame corresponding to an equivalent single long integration time image (of $\sim5$~s duration). From this, a well calibrated Strehl ratio calculator was used to extract the Strehl ratio. Using Marechal's approximation for the Strehl ratio given by
\begin{equation}\label{eq:strehl}
SR\sim\frac{1}{e^{(2\pi\omega)^{2}}},
\end{equation}
where $\omega$ is the normalised wavefront error at a given wavelength, the amount of RMS wavefront error was determined. The RMS wavefront error was calculated at a wavelength of $1550$~nm.

\subsection{Coupling into a SMF versus RMS wavefront error}\label{coupvrms}
\begin{figure}[ht!]
\centering 
\includegraphics[width=0.99\linewidth]{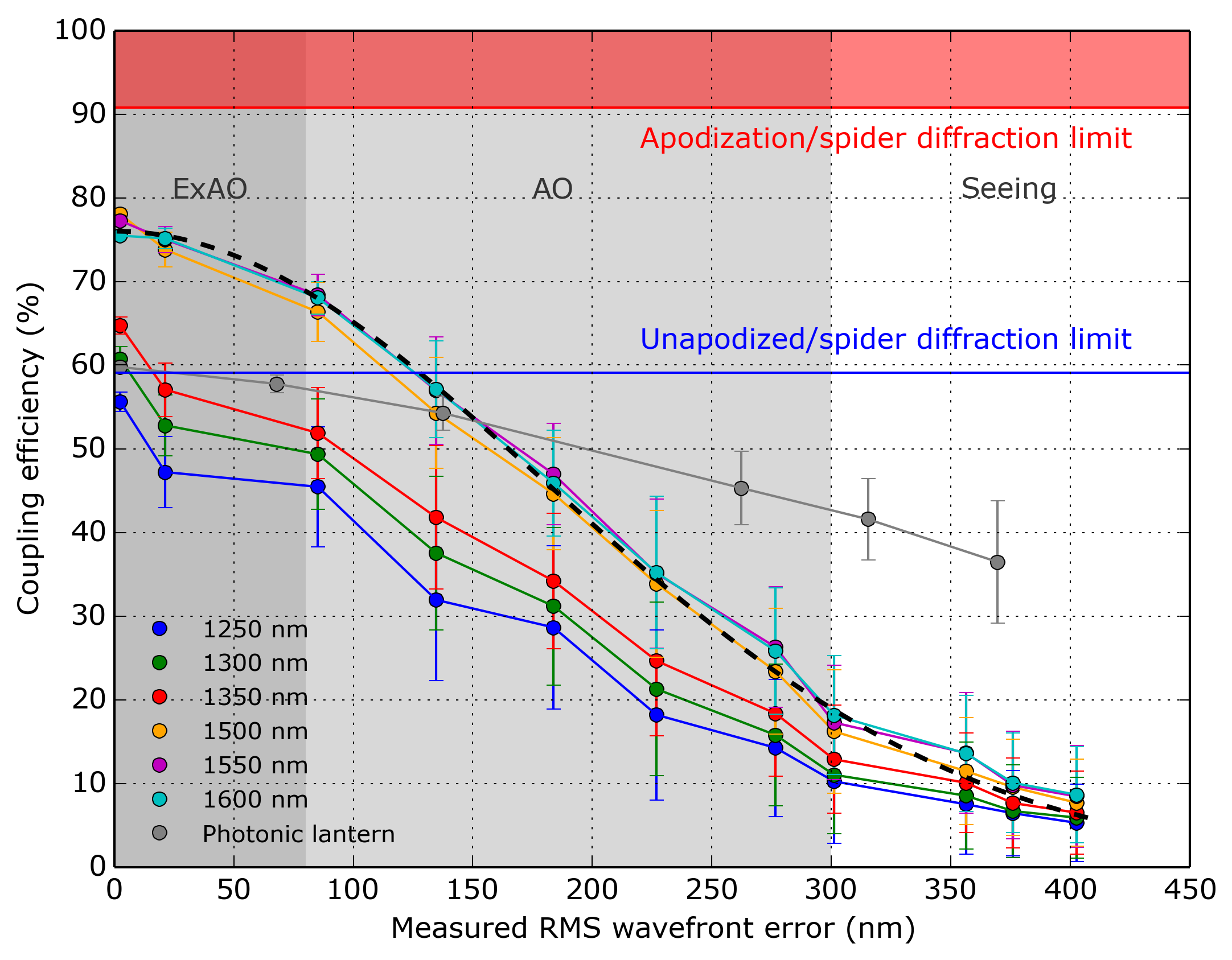}
\caption{\footnotesize Coupling efficiency as a function of calculated RMS wavefront error at $1550$~nm. The ExAO, AO and seeing regimes are highlighted for convenience. The coupling efficiency limit due to spiders and apodisation optics is highlighted in red, while the limit without apodisation is in blue. The dashed black line is Marechal's approximation to the Strehl based on the wavefront error calculated from Eq.~\ref{eq:strehl}. A clear correlation between the Strehl and coupling is visible.}
\label{fig:coupvsrms}
\end{figure}

The coupling efficiency as a function of the RMS wavefront error is shown in Fig.~\ref{fig:coupvsrms}. This way of presenting the data is instrumental for people who build AO systems and think in units of RMS wavefront error.

Overlaid in Fig.~\ref{fig:coupvsrms} is Marechal's approximation to the Strehl ratio given the RMS wavefront error (black dashed line). Unsurprisingly this is a good fit to the coupling measurements as this equation was used to convert the Strehl ratio to wavefront error in the first place, but nonetheless reassuring to confirm. In the ExAO regime, where the wavefront error is $<80$~nm RMS or that is the wavefront flatness is $<\lambda/19$ at $1550$~nm, a coupling $>67\%$ can be achieved as outlined above. For standard AO levels of correction, where the wavefront error is $<300$~nm RMS or that is the wavefront flatness is $<\lambda/5$ at $1550$~nm, a coupling $>18\%$ can be achieved. There is clearly an advantage to exploiting an ExAO system for injecting into a SMF, however, for the brightest targets it is possible to use a conventional AO system for this application as well. Indeed, in median seeing where a conventional AO system can achieve $200$~nm RMS wavefront error the coupling could be as high as $\sim40\%$ if PIAA optics are used. By doing so this opens the possibility of employing these techniques on many more telescopes.

\subsection{Preliminary on-sky data}\label{apponsky}
Alpha Bootis (Arcturus, K$0$ spectral type, R-mag$=-1.03$, H-mag$=-2.81$) was observed on the night of the $18^{th}$ of March. The seeing varied between $0.6"$ and $1.0"$ in V-band throughout the night. The average Strehl ratio was measured immediately following data collection with the photometer and determined to be $46.3\pm2.1\%$ in H-band. We note that the Strehl ratio could not be measured during data acquisition with the photometer. The same alignment process was used as described for the observations on the $21^{st}$ of March.

\begin{figure}[!t]
\centering 
\includegraphics[width=0.99\linewidth]{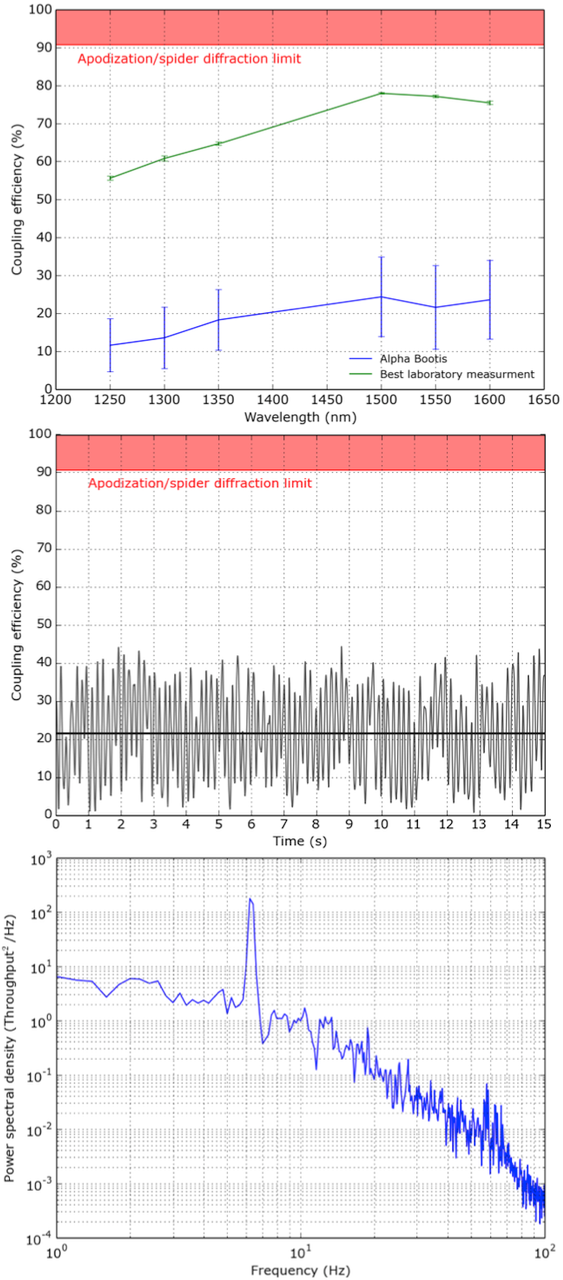}
\caption{\footnotesize (Top) Coupling as a function of wavelength (target: Alpha Bootis (Arcturus) taken
on the night of the $18^{th}$ of March). The green line shows the best laboratory result obtained with no wavefront error applied. (Middle) Time series of the $1550$~nm spectral channel. The dark solid line shows the mean over the $15$~s data-collection period. (Bottom) A PSD of the time series in the middle panel.}
\label{fig:onskyalphboo}
\end{figure}

\begin{figure}
\centering 
\includegraphics[width=0.99\linewidth]{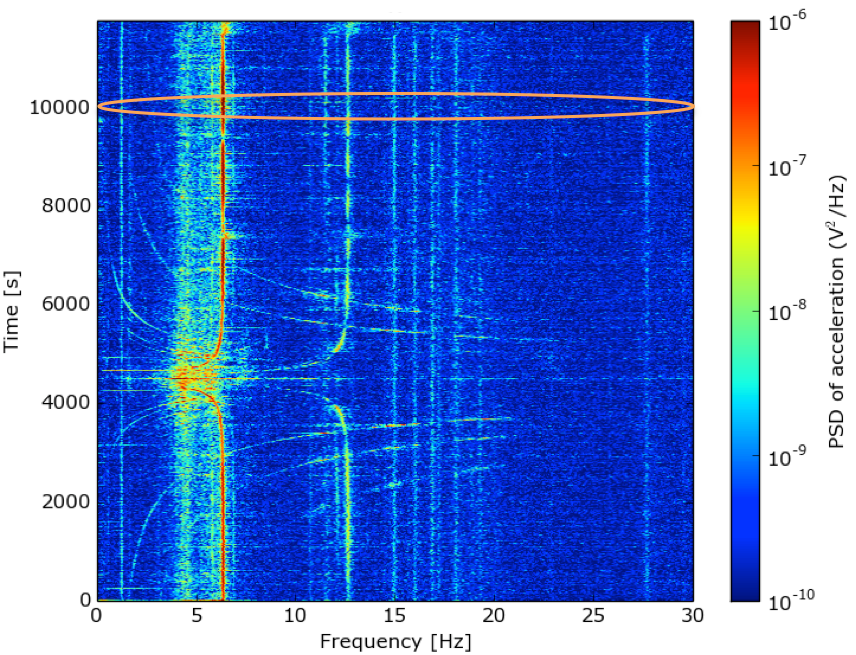}
\caption{\footnotesize Data from accelerometers attached to the top ring of the telescope. The data is only presented for the elevation axis. The orange ellipse indicates the period when the data was acquired with the photometer.}
\label{fig:accel}
\end{figure}

The coupling efficiency as a function of wavelength is shown in the upper panel of Fig.~\ref{fig:onskyalphboo}.  
It can be seen that the coupling reaches $24\%$ between $1500$ and $1600$~nm. The error bars are much larger than those taken in the laboratory indicating a huge fluctuation in the coupling as a function of time. The middle panel of Fig.~\ref{fig:onskyalphboo} shows a time series of the coupling efficiency for the $1550$~nm spectral channel. The coupling efficiency seems to indeed be heavily modulated, so much so that at some points the coupling is almost zero and at others it is as high as $40\%$. The dark black line indicates the mean value of the time series which represents the data point at $1550$~nm in the top panel. The line is positioned halfway up the data set, which is consistent with a strong modulation of the flux. A power spectral density (PSD) plot is offered in the lower panel of Fig.~\ref{fig:onskyalphboo} which clearly shows that a strong periodic modulation is present at $6.3$~Hz. Otherwise the spectrum is near featureless. Accelerometers attached to the top ring of the telescope reveal the source of this vibration. Figure~\ref{fig:accel} shows the PSD as a function of time for the accelerometer for the elevation axis of the telescope.  
A strong resonance can clearly be seen at $6.3$~Hz consistent with the results from the photometer. These are known telescope vibrations that manifest as tip/tilt jitter in the focal plane of the detectors and fibre injection and cause the coupling to oscillate~\citep{lozi2016}. These vibrations are persistent despite both the AO$188$ and PyWFS control loops operating. During this observation we did not have a working LQG or an operational LOWFS. Both of which would help in the future for mitigating these vibrations. The interesting feature here is the amplitude of the vibration. The fact that the coupling was modulated by almost $100\%$ would indicate that either the PSF was optimally aligned but had a peak-to-peak motion of the order of $3$ to $4~\lambda/D$ (i.e. $1.5$--$2$ full width at half-maximums in radius), or that the fibre was not optimally aligned in the first instance and the peak-to-peak modulation was $1$ to $2~\lambda/D$, post AO correction. Interestingly, the measured Strehl ratio of $\sim46\%$ in H-band should deliver a coupling efficiency at $1550$~nm of $\sim36\%$ from Fig.~\ref{fig:coupvssr}. Indeed, if one were to remove the modulation due to vibrations, the coupling efficiency seems to peak around this value (middle panel in Fig.~\ref{fig:onskyalphboo}), highlighting that results in Fig.~\ref{fig:coupvssr} can be used to infer the expected coupling from the Strehl ratio.

\end{document}